\documentclass[useAMS,a4,usenatbib,preprint]{mn2e}

\usepackage{graphicx}
\usepackage{color}
\usepackage{amsmath,environ}
\usepackage{subfigure}
\usepackage{epsfig,epstopdf}
\usepackage{hyperref,ulem}
\usepackage[T1]{fontenc} 
\usepackage{aecompl}
\newcommand{\be}{\begin{equation}}
\newcommand{\ee}{\end{equation}}
\newcommand{\bea}{\begin{eqnarray}}
\newcommand{\eea}{\end{eqnarray}}

\newcommand{\kpc}{\rm kpc}
\newcommand{\pc}{\rm pc}
\newcommand{\lsim}{\mathrel{\mathop{\kern 0pt \rlap
  {\raise.2ex\hbox{$<$}}}
  \lower.9ex\hbox{\kern-.190em $\sim$}}}
\newcommand{\gsim}{\mathrel{\mathop{\kern 0pt \rlap
  {\raise.2ex\hbox{$>$}}}
  \lower.9ex\hbox{\kern-.190em $\sim$}}}

\title[$\gamma$-ray anisotropies]{$\gamma$-ray anisotropies from dark matter in the Milky Way:
the role of the radial distribution \footnote{LAPTH-009/14, PCCF RI 14-03}}

\author[Calore et al.]{F. Calore$^{1}$, V. De Romeri$^{2}$, M. Di Mauro$^{3,4,5}$, F. Donato$^{3,4}$, J. Herpich$^{6}$, A.V. Macci\`{o}$^{6}$, 
\newauthor{L. Maccione$^{7,8}$}
\vspace*{6pt}\\
$^{1}$ GRAPPA Institute, University of Amsterdam, Science Park 904, 1090 GL Amsterdam, The Netherlands\\
$^{2}$ Laboratoire de Physique Corpusculaire, CNRS/IN2P3 - UMR 6533, Campus des C\'ezeaux, 24 Av. des Landais, F-63177 Aubi\`ere Cedex, France \\
$^{3}$ Dipartimento di Fisica, Universit\`a di Torino,  via Giuria 1, I-10125 Torino, Italy\\
$^{4}$ INFN, Sezione di  Torino, I-10125 Torino, Italy\\
$^{5}$ Laboratoire d'Annecy-le-Vieux de Physique Th\'eorique (LAPTh), Univ. de
Savoie, CNRS, B.P.110, Annecy-le-Vieux F-74941, France\\
$^{6}$ Max-Planck-Institut f\"ur Astronomie, K\"onigstuhl 17, 69117 Heidelberg, Germany\\
$^{7}$ Ludwig-Maximilians-Universit\"{a}t, Theresienstra{\ss}e 37, D-80333 M\"{u}nchen, Germany and \\
$^{8}$ Max-Planck-Institut f\"{u}r Physik (Werner Heisenberg Institut), F\"{o}hringer Ring 6, D-80805 M\"{u}nchen, Germany}

\begin{document}
\date{LAPTH-009/14, PCCF RI 14-03}

\maketitle
\label{firstpage}

\begin{abstract}
The annihilation of dark matter particles  in the  halo of galaxies may end up into  $\gamma$-rays, which travel almost
unperturbed till to their detection at Earth. 
This annihilation signal can exhibit an anisotropic behavior quantified by the angular power spectrum, whose
properties strongly depend on the dark matter  distribution and its clumpiness.
We use high resolution pure dark matter N-body simulations to quantify the contribution of different components (main halo and satellites) 
to the global signal as a function of the analytical profile adopted to describe the numerical results.
We find that the smooth main halo dominates the angular power spectrum of the $\gamma$-ray signal 
up to quite large multipoles, where the sub-haloes 
anisotropy signal starts to emerge, but the transition multipole strongly depends on the assumed radial profile.  
The extrapolation down to radii not resolved by current numerical simulations can affect both the normalization $and$ 
the shape of the $\gamma$-ray  angular power spectrum.
For the sub-haloes described by an asymptotically cored dark matter distribution, the angular power spectrum 
shows an overall smaller normalization and a flattening at high multipoles. 
Our results show the criticality of the dark matter density profile shape in $\gamma$-ray anisotropy searches, and 
evaluate quantitatively the intrinsic errors occurring when  extrapolating the dark matter radial profiles
down to spatial scales not yet explored by numerical simulations.
\end{abstract}


\section{Introduction}
\label{sec:intro} 

One of the most reliable solutions to the missing mass in the Universe implies 
that it is constituted by  Weakly Interacting Massive Particles (WIMPs), 
clustered in galaxies as dark haloes.
The astrophysical evidence of these particle dark matter (DM) candidates can 
 be explored by direct as well as indirect detection techniques. 
In the latter case, the idea is that  WIMP DM may annihilate in pairs and produce charged particles 
and $\gamma$-rays, detectable as rare components in cosmic rays. 
Differently from the charged cosmic rays, the flux of $\gamma$-rays arriving at the 
Earth is not deflected by  magnetic fields and traces back directly to its sources. 
The search for DM through $\gamma$-rays is therefore a preferential tool for exploiting the 
properties of its spatial distribution (\cite{2012PDU.....1..194B} and refs. therein).  

The excellent performances of the Large Area Telescope (LAT) on the  Fermi $\gamma$-ray space Telescope ({\it Fermi}) 
have let the exploration for a DM component in the Milky Way, in extragalactic nearby objects, as well as 
in cosmological structures \citep{2012ApJ...761...91A,2010ApJ...712..147A,2010JCAP...05..025A,Abdo:2010dk, 2011PhRvL.107x1302A}. 
At high galactic latitudes, a faint $\gamma$-ray irreducible emission has been measured,
and shown to be isotropic at a high degree \citep{Ackermann:2012uf}. 
The {\it Fermi}-LAT has already reported the detection of a non-zero angular power spectrum (APS) above the noise 
level in the multipole range $\ell \sim 155\div504$, corresponding to an angular scale $\lsim 2^\circ$, with a 
significance ranging from $5.3\sigma$ between 2 and 5 GeV to $0.8\sigma$ between 10.4 and 50 GeV \citep{Ackermann:2012uf}.

Different predictions for the APS have been proposed for various populations of unresolved sources, both of astrophysical 
\citep{2012PhRvD..86f3004C,SiegalGaskins:2010mp,2012JCAP...11..026H} and of DM origin 
\citep{Fornasa09,2013MNRAS.429.1529F,2013PhRvD..87l3539A,SiegalGaskins:2008ge}. 
Since it is expected that the statistical properties of the DM distribution in galactic and 
extragalactic space are different from those of 
standard astrophysical objects, the study of the APS ascribable to DM sources may be an important signature worth to be explored.

The intensity of the $\gamma$-ray signal depends on a particle physics term - describing the strength and the 
energy spectrum of the annihilation - and on the  DM density in collapsed structures. 
While the first factor includes the details of 
the assumed particle physics  model for the WIMPs, 
the second one has an astrophysical origin and it is usually modeled according to the results of 
 cosmological collision-less simulations, generally predicting a steepening of the DM profile in the inner parts of the resolved halos (e.g. \cite{aquarius,kuhlen08} and references therein).  Therefore, the most likely detectable
targets have  been identified with especially dense regions such as 
the galactic center \citep{Gomez-Vargas:2013bea,Fermi-LAT:2013uma}, 
and the center of any DM sub-structure orbiting in the Milky Way halo, 
like faint and ultra faint dwarf galaxies \citep{Walker:2011fs,Ackermann:2013yva}. 

The APS gives the measure of a signal correlation between two angular scales, and, in turns, between two spatial scales.
For a source located at the galactic center, like the main halo of the Milky Way, the APS at multipoles, for example, 
$\ell>500$ probes the DM 
distribution at $R < \pi/500\cdot8.5~\kpc \sim 40 ~\pc$.  The study of the APS at 
$l\gsim 500$ requires therefore to know the DM profile at scales much below the resolution ($\sim$200 pc) of current state of the art numerical simulations for structure formation
\citep{aquarius,kuhlen08,2009MNRAS.398L..21S}.
Several profile parameterizations provide excellent fits to the DM distribution of simulated halos 
\citep{NFW,2004MNRAS.349.1039N,2006AJ....132.2701G,2008MNRAS.391.1940M,2009MNRAS.398L..21S}. 
However, when extrapolated below the resolution
limit of cosmological simulations, different profiles predict very different central densities.

In this paper we discuss in detail these points when applied to the anisotropy in the $\gamma$-ray flux from DM annihilation, namely:
i) the intrinsic uncertainty due to the extrapolation to short distances of the DM distribution determined from numerical simulations; 
ii) the different signatures in the APS in connection with the various density profiles (cored and cuspy).

\bigskip 
\section{The $\gamma$-ray flux from DM annihilation}
\label{sec:DMflux}  
The $\gamma$-ray flux $\mathrm{d} \Phi_{\gamma}/\mathrm{d} E_{\gamma}$
 from DM annihilating particles is defined as the number of photons collected by a detector per unit of time, area, solid angle
and observed energy $E_{\gamma}$. When looking at the direction $\psi$ and $\theta$ 
(longitude and latitude in Galactic coordinates, respectively) in the sky, 
by an experiment with spatial resolution $\alpha$ and under a solid angle $\Delta \Omega= 2\pi\,(1-\cos\,\alpha)$,
 it may be expressed as: 
\bea
\label{eq:dPhidE}
\frac{\mathrm{d} \Phi_{\gamma}}{\mathrm{d} E_{\gamma}}(E_{\gamma},\psi,\theta,\Delta
\Omega) =\\ \nonumber 
 \frac{1}{4\pi}\, \frac{\langle\sigma_{\rm ann} v\rangle}{2m_{\chi}^{2}}
  \cdot \sum_{i} B_{i} \cdot  \frac{dN^{i}_{\gamma}}{dE_{\gamma}}\, 
   \int_{0}^{\Delta \Omega}\, \mathrm{d} \Omega\int_{\rm{l.o.s}} \rho^{2}(r(s,\psi,\theta)) ds.
   \eea

Here $m_\chi$ is the mass of the DM particle and $\langle\sigma_{\rm ann}v\rangle$ is the annihilation cross section times the relative velocity 
averaged over the DM velocity distribution. $B_i$ is the branching ratio into the final state $i$ and $dN^{i}_{\gamma}(E_\gamma) / dE_{\gamma}$ 
is the photon spectrum per annihilation (which depends on the annihilation channels). 
The sum is in principle performed over all the annihilation channels.
The last term in Eq. \ref{eq:dPhidE} contains the (squared) DM density $\rho(r)$ ($r$ being 
the galactocentric distance) integrated along a distance $s$ from the Earth in the direction along the line of sight (l.o.s),
and  in the observational cone of solid angle $\Delta\Omega$. 
In the following of our analysis, if not differently stated, we will choose as representative the annihilation into the $\bar{b}b$ quark channel
with $B_{\bar{b}b}$=1, and fix $m_\chi$=200 GeV, $E_\gamma$=4 GeV and 
$\langle\sigma_{\rm ann} v\rangle = 3\cdot 10^{-26}$ cm$^3$ s$^{-1}$. 
We remark that this choice does not affect our main results, since throughout the 
analysis the particle physics factor may be considered as a mere normalization of the APS.

\subsection{Simulations for the  DM spatial distribution}
\label{sec:rho}

The simulations presented in this paper are the pure DM N-body counterparts of the MaGICC
(Making Galaxies in a Cosmological Context) simulations suite (\cite{2013MNRAS.428..129S,2014MNRAS.437..415D} for more details). 
The galaxy we discuss in details is g15784, which has a virial 
mass of $1.48 \times 10^{12} \rm M_\odot$, very close to the mass of the Milky Way \citep{2008ApJ...684.1143X}. 
We resolve a total of 27 substructures in the simulation in a mass range of $10^{8.6}-10^{9.6} M_\odot$.

For determining the $\gamma$-ray emission, as clear from Eq. \ref{eq:dPhidE}, a special role is deserved to the radial density profile of the
DM halo $\rho(r)$, with particular attention to the central region. This is true both for the central
smooth halo as well as for the sub-structures.

We have decided to use three different analytical profiles to describe the DM distribution in our simulation:
the widely used Navarro, Frenk \& White profile (\citet{1997ApJ...490..493N}, NFW hereafter), 
the Einasto profile (\citet{1965TrAlm...5...87E,1968PTarO..36..341K,1969Afz.....5..137E}, Ein hereafter) which
has been shown to be a better representation of the DM distribution in simulated
haloes (\citet{duttonmaccio}), and the profile suggested by Moore and Stadel
(\citet{2009MNRAS.398L..21S}, MS hereafter): %
\begin{eqnarray}
\label{eq:profileNFW} \rho(r) &=& \rho_{0} \left[ \left( \frac{r}{R_c} \right) \left(1+ 
 \frac{r}{R_c} \right)^2 \right]^{-1}   \qquad {\rm (NFW)} \\
\label{eq:profileEinasto} \rho(r) &=& \rho_{0}\exp\left(-\frac{2}{\alpha_{E}}\left[\left(\frac{r}{R_{s}}\right)^{\alpha_{E}}-1\right]\right)\qquad   {\rm (Ein)} \\
\label{eq:profileMS} \rho(r) &=& \rho_{0}\exp\left(-\lambda\left[\ln\left(1+\frac{r}{R_{\lambda}}\right)\right]^{2}\right)  \qquad {\rm (MS)}\;
\end{eqnarray}
where $\rho_{0}, R_c, R_{s}, \alpha_{E},\lambda, R_{\lambda} $ are the free parameters in the different
analytic profiles.
In this latest parameterization the density profile is linear down 
to a scale $R_{\lambda}$, beyond which it approaches  the central maximum density $\rho_0$ as $r \rightarrow 0$. 
This fitting function is extremely flexible and makes possible to reproduce at the same time both cuspy 
and cored profiles (e.g. \citet{2012MNRAS.424.1105M}).
\begin{figure*}
\begin{center}
\includegraphics[width=0.48\textwidth]{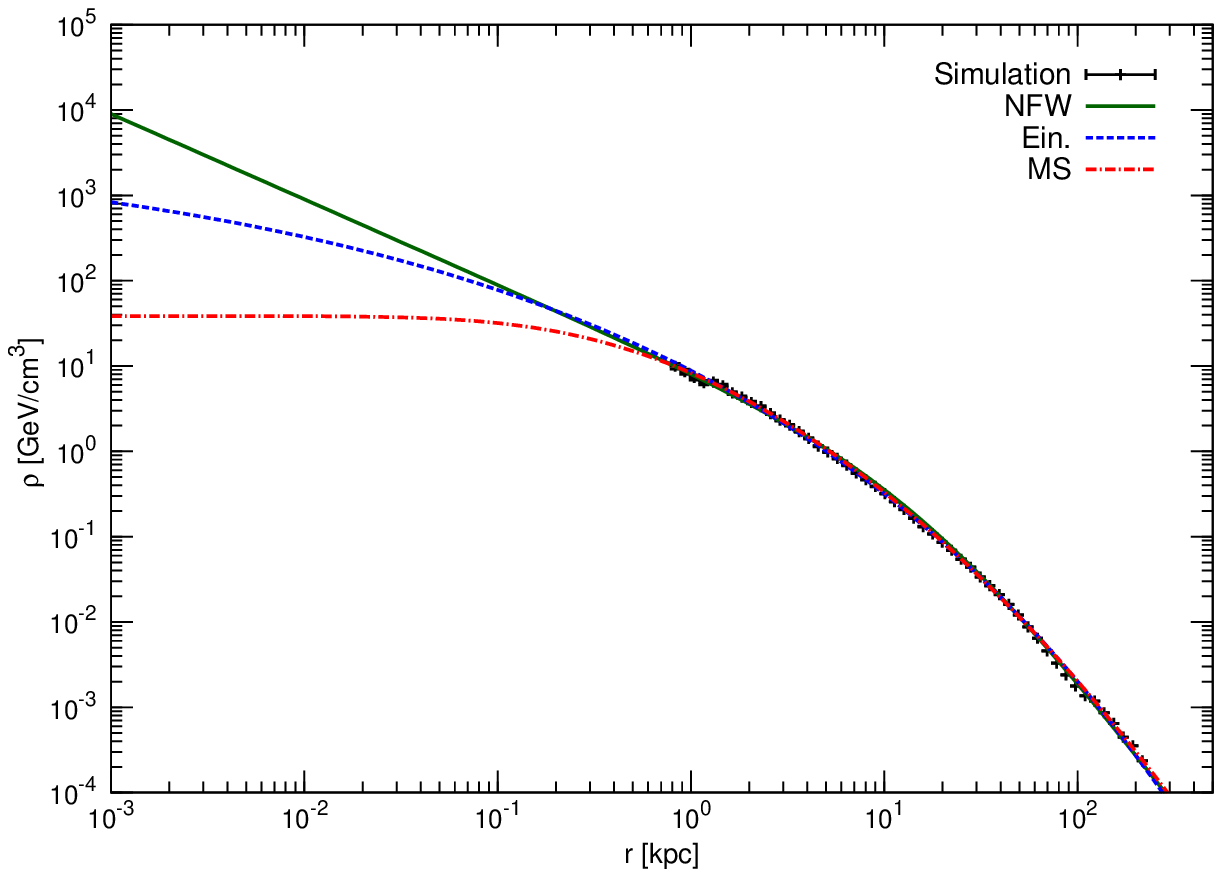}
\includegraphics[width=0.48\textwidth]{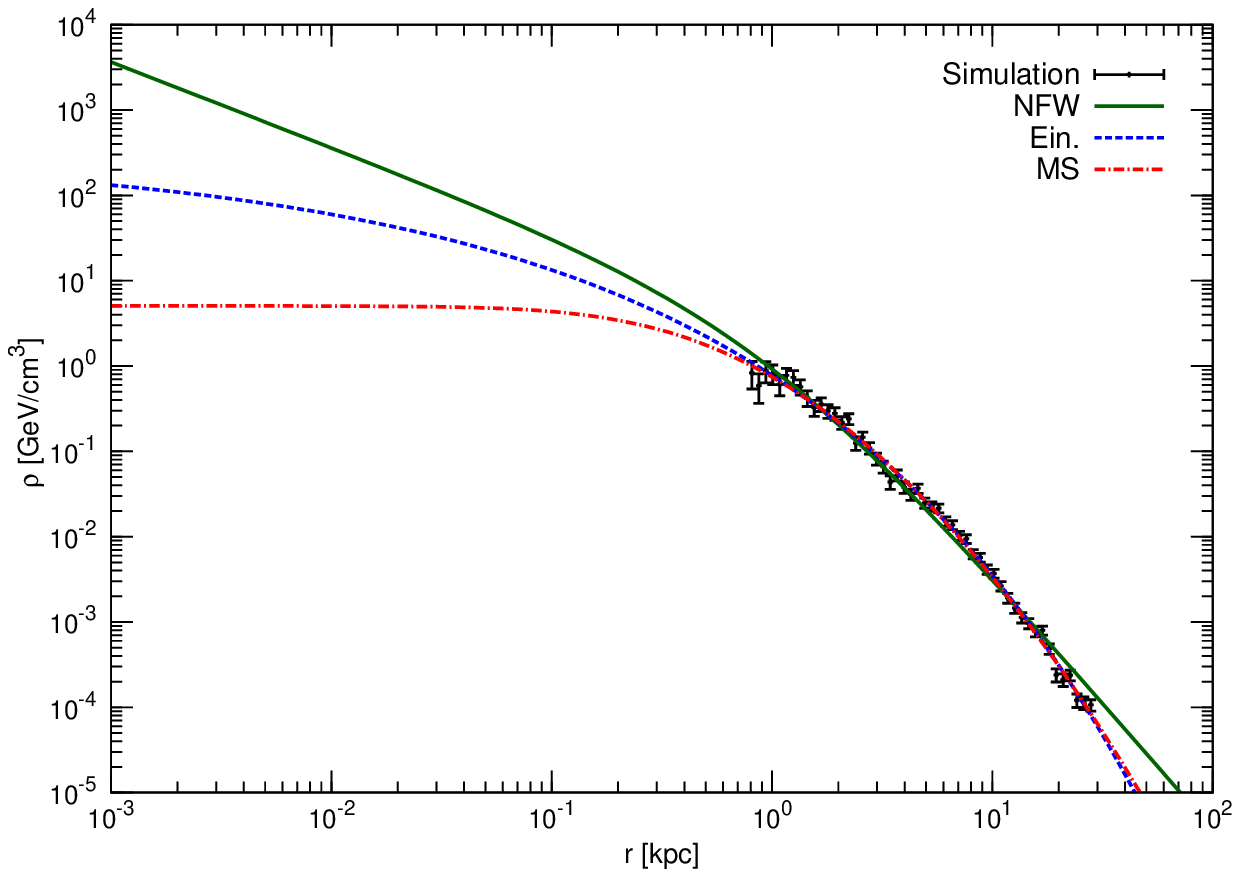}
\caption{DM density profile as a function of the radius $r$ from the center of the halo. 
The left panel displays the results for the main halo, the right panel for a  sub-halo resolved in the simulation g15784.
 The simulation data are shown by black dots, while the different
profiles described in Eqs. \ref{eq:profileNFW} (NFW), \ref{eq:profileEinasto} (Ein), \ref{eq:profileMS} (MS),
and their extrapolation to smaller scales, are shown by the red solid,  blue dotted and green dot-dashed lines, respectively.}
\label{fig:profiles}
\end{center}
\end{figure*}
The results of the different fits to the DM distribution are shown in  Fig.~\ref{fig:profiles}. 
We notice that throughout this work the main halo is intended to be the total DM halo.
In the left panel, it is clear that all the different profiles described above provide a very good fit to the numerical radial density  
on the whole range probed by the simulation (0.8-250 kpc). On the other hand, they dramatically diverge
when extrapolated beyond the resolution limit of the simulation. The MS profile
predicts an extended core below 50-100 pc, while the Einasto and NFW profiles both imply an increased density 
towards the center, even though with a quite different slope. 
As a result, the central DM halo density at the $\approx10$ pc scales - the most relevant scale for $\gamma$-rays
production - differs by a factor of fifty between the two most 
extreme cases (MS and NFW) and by an order of magnitude between Einasto and MS radial profiles. 
Similar results may be 
drawn for a sub-halo resolved in the simulation g15784 and fitted with the same functions 
(Fig.~\ref{fig:profiles}, right panel). In this case, 
the NFW density profile shows some tension also with data at larger radii. 
The central DM halo density at about 10 pc differs by more than two orders of magnitude between the two most 
extreme cases (MS and NFW) and by an order of magnitude between the two cuspy profiles
(Einasto and NFW).

These simple plots show how problematic (and dangerous) it is the extrapolation of cosmological
N-body simulations results on very small spatial scales. As we will see in the rest of the paper,
this extrapolation has profound effects on the predicted  $\gamma$-ray DM signal and the relative
contribution of different components like the central halo and its satellites  to the APS.

\subsection{The angular power spectrum of $\gamma$-ray anisotropies.}
\label{sec:anis}
The  intensity APS $C_{\ell}$ of a map $I(\Psi)$, where $\Psi$ is a direction in the sky,
is given by the coefficients 
\be 
C_{\ell} = \frac{1}{2\ell+1} \sum_{\mid m \mid < \ell}  |a_{\ell m}|^{2} ,
\label{eq:Cl_Int}
\ee
with the $a_{\ell m}$ determined by expanding the sky map in spherical harmonics, after subtracting the average value of the intensity 
over the region of the sky considered: 
\begin{equation}
I(\Psi)= \frac{\mathrm{d} \Phi}{\mathrm{d} E}(\Psi) - \langle \frac{\mathrm{d} \Phi}{\mathrm{d} E}(\Psi) 
\rangle = \sum_{\ell=0}^\infty \sum_{m=-\ell}^{m=\ell}  a_{\ell m} Y_{\ell m}(\Psi).
\label{eq:intAPS}
\end{equation}
The $\gamma$-ray intensity maps and their power spectra have been generated by using the
HEALPix software \citep{2005ApJ...622..759G}.
Depending on the parameter order $k$, the number of pixels of the map is $N_{\rm pixel}=12\cdot2^{2k}$.
Hence, the solid angle of one pixel of the map is $\Delta \Omega = 4\pi / N_{\rm pixel}$. 
We fix k=13, so that $ \Delta \Omega  = 1.56\cdot10^{-8}$ sr
for a  corresponding  scale of about $1$ pc, except for 
the results of the Monte Carlo simulation where the order parameter  is fixed to k=9 for $\Delta \Omega  = 4\cdot10^{-6}$ sr.
The maximum  multipole number $l_{\rm max}$ compatible with a fixed map resolution 
is $l_{\rm max} \sim 2\cdot2^{k}$, therefore $\sim 1.6\cdot 10^4$ (1024) for k=13 (k=9) \citep{2005ApJ...622..759G}.

\section{Results}
\label{sec:results}

\begin{figure*}
\includegraphics[scale=0.35]{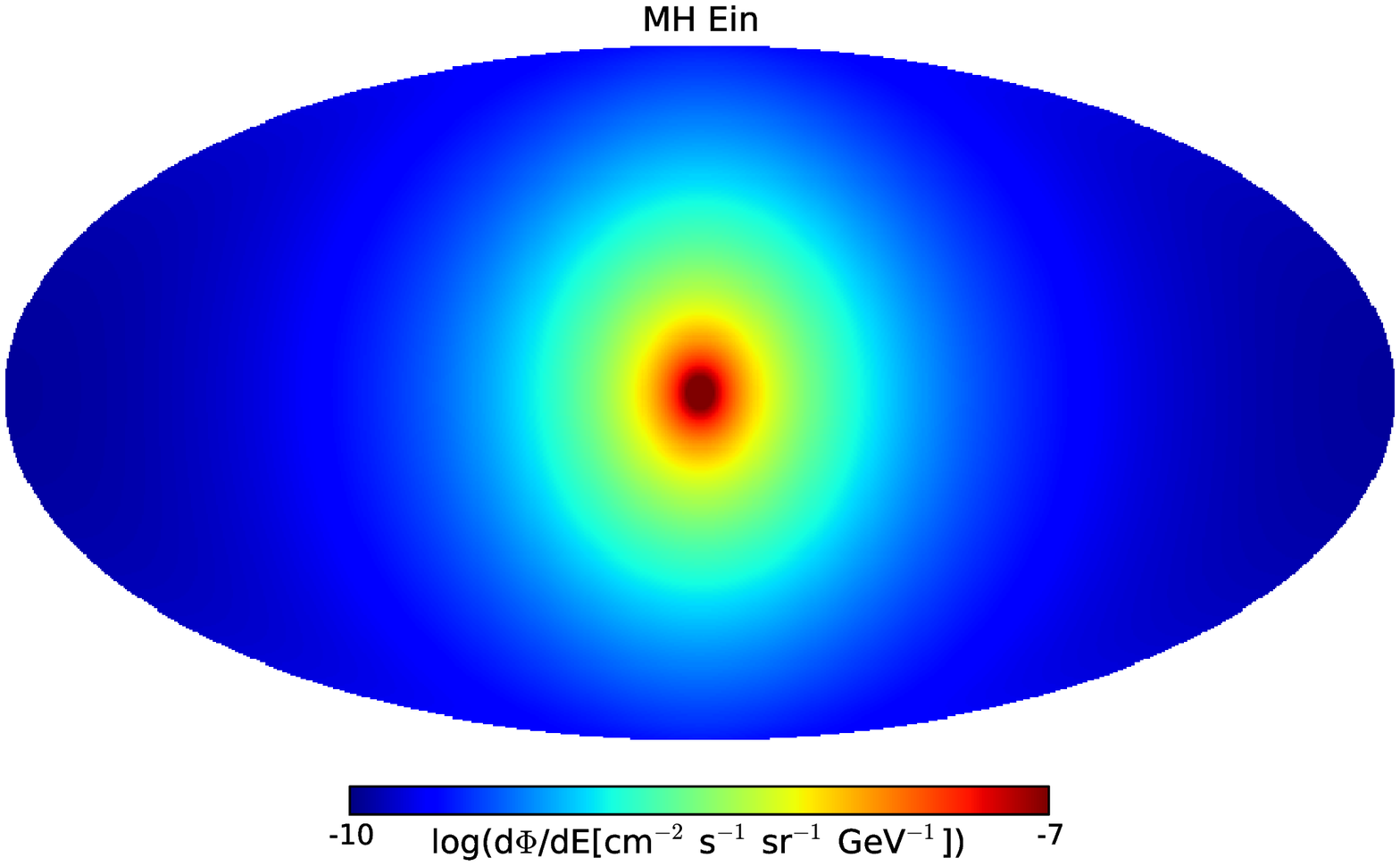}
\includegraphics[scale=0.35]{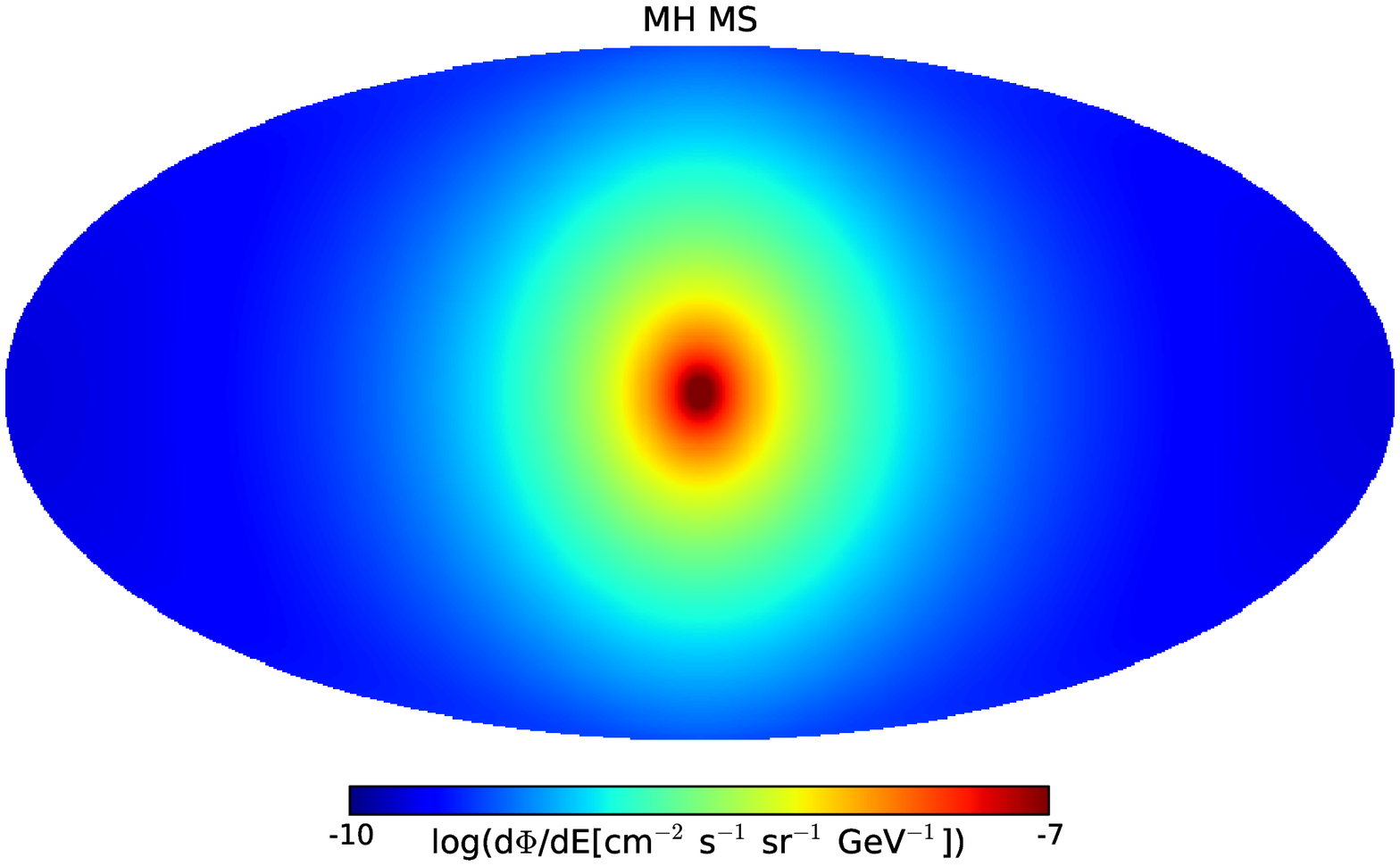}
\includegraphics[scale=0.35]{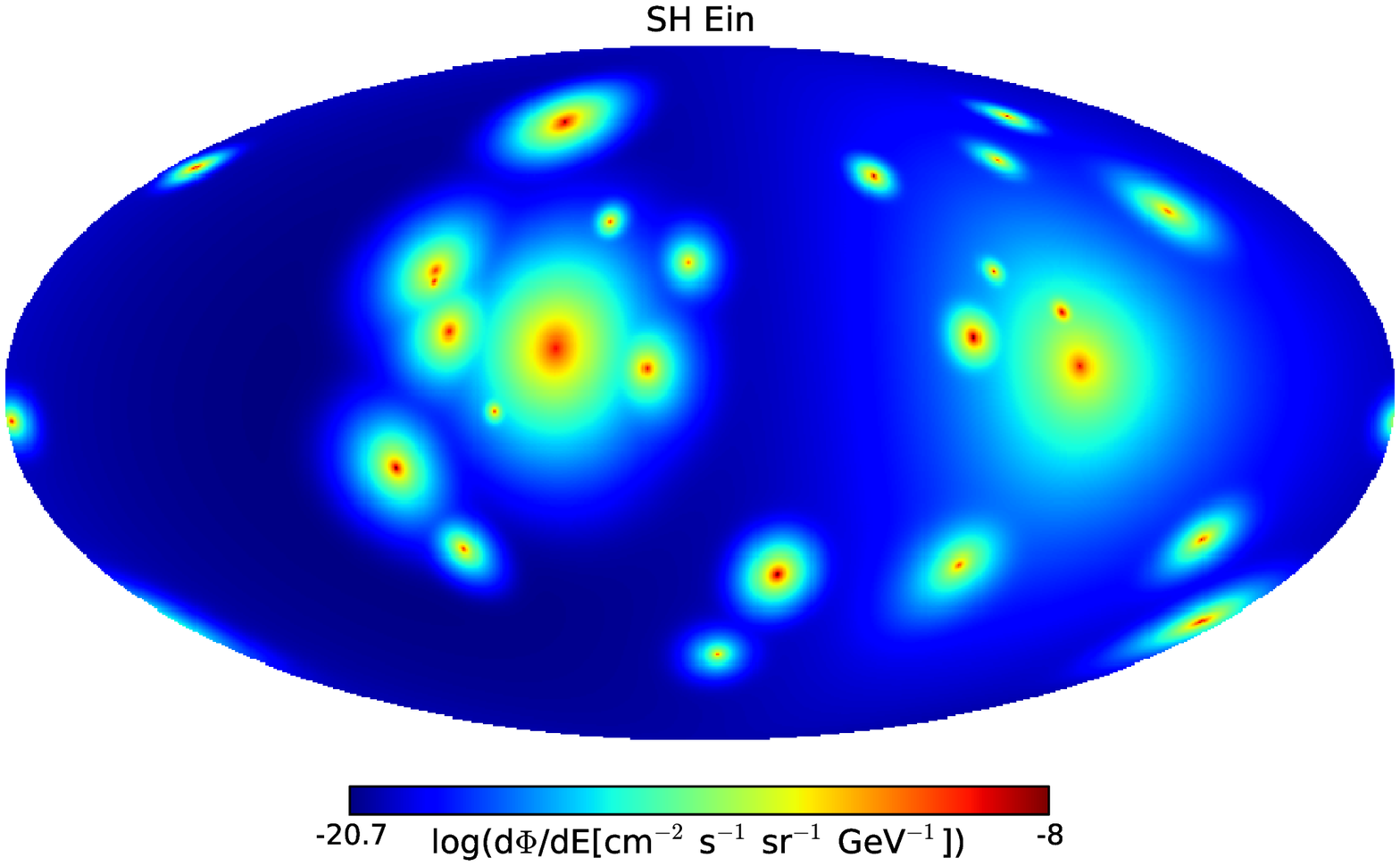}
\includegraphics[scale=0.35]{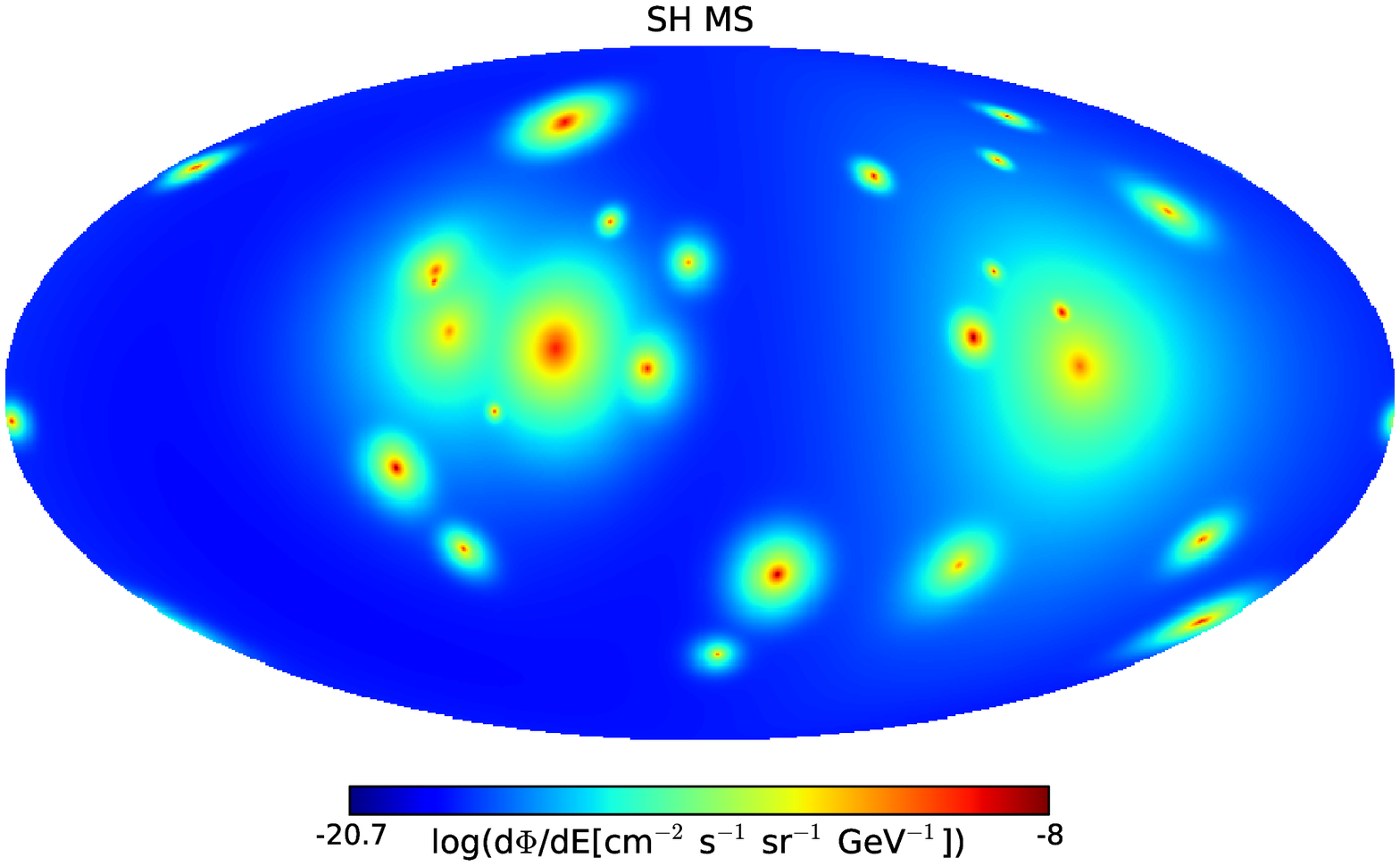}
\caption{All-sky maps of the $\gamma$-ray emission at $E_\gamma$=4 GeV, from the
annihilation of  $m_\chi$= 200 GeV DM in the simulated galactic halo g15784.  
In the left (right) panels the main smooth halo (MH, upper plot) and sub-haloes (SH, lower plots) are interpreted with the Einasto (MS) 
radial density profile, Eq. \ref{eq:profileEinasto} (Eq. \ref{eq:profileMS}).  }
\label{fig:map}
\end{figure*}
We have computed the space distribution of the $\gamma$-ray emission from DM annihilation  based on the g15784 
halo simulation, described in Sect. \ref{sec:rho}. 
The resulting simulated sky is illustrated in Fig. \ref{fig:map}, where we plot the $\gamma$-ray emission maps at 
$E_\gamma$=4 GeV, from a DM halo composed by WIMPs 
with $m_\chi$= 200 GeV (see Sect. \ref{sec:DMflux}  for details).  
 In the left panels, the main halo and the sub-haloes are interpreted with the Einasto DM
spatial distribution in Eq. \ref{eq:profileEinasto}, 
while the right panels show the same halo when described by the MS 
 $\rho(r)$ (Eq. \ref{eq:profileMS}).  The color code
refers to the intensity of the map and goes from blue to red with increasing flux, 
with different scales for the main halo and the sub-structures.
The emission from the smooth component (top panels), as expected, extends at larger radii for the
MS parameterization, while it is more concentrated in the center for the Einasto
profile, given the steeper behavior towards the center of the galaxy. 
The same argument applies to each sub-halo of the simulation: when interpreted as
distributed according to the Einasto profile, the most of the $\gamma$-ray emission
of each sub-halo mainly originates from the very center of the sub-structure, while in the
case of MS the emission is distributed over a larger region, rightly  because the core of the DM 
profile is more extended. 

\begin{figure}
\includegraphics[width=0.48\textwidth]{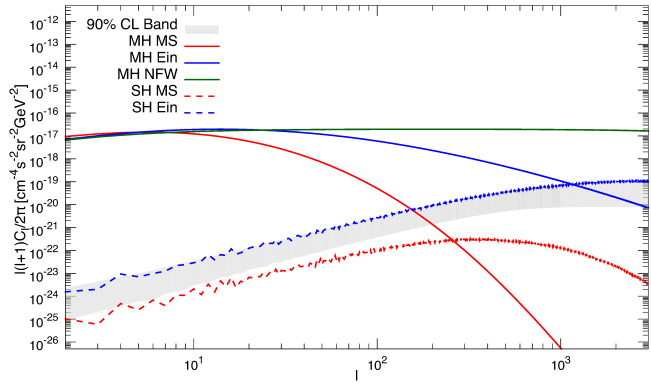}
\caption{Intensity APS for the simulated halo g15784, as a function of the multipole number. The solid blue (red) line describes the smooth halo
according to the Einasto (MS) profile, while the dotted lines are for the corresponding sub-haloes contribution.
The green solid line displays results for an NFW profile fitting the smooth halo. The grey band refers to the 90\% CL uncertainty related to the orientation of the sub-haloes of the g15784 simulation.}
\label{fig:cl}
\end{figure}

We have calculated  the intensity APS for the all-sky $\gamma$-ray maps of the simulated galaxy, for both the smooth halo
and the resolved sub-structures. 
The $\gamma$-ray intensity in a given direction is obtained by piling up the contribution from all 
sub-haloes encountered along the l.o.s., up to a distance of 500 kpc. 
The results are shown in Fig. \ref{fig:cl}, where the intensity APS for
the main halo and the sub-haloes is described, alternatively, 
 by the parameterizations of Eqs.~\ref{eq:profileEinasto} and \ref{eq:profileMS}. 
The figure has been obtained setting the HEALPix resolution k=13.  
The halo, when interpreted in terms of the peaked Einasto profile, yields much more power at small radial scales (high $l$), 
and this is true for both the smooth halo and the sub-haloes. 
The two profiles give comparable APS only for $l\lsim$ 10, while at $l$=100 the Einasto APS is about 
two orders of magnitude higher than the MS one. At very small scales, such as $l$=1000
or, equivalently $\sim 30$ pc, the main halo within the MS profile does not contribute any longer to the 
anisotropy of the sky, while the Einasto profile still provides a sizable APS (about eight orders of magnitude above the MS contribution). 
For illustrative purposes, we also plot the intensity APS for the main smooth halo interpreted in terms of a NFW 
$\rho(r)$ (which indeed fails to properly fit the simulated sub-haloes). The implied APS is very high at all scales, even with respect 
to the Einasto modeling. At $l$=1000, the cuspiness of the NFW profile 
gives an APS 100 times more intense than for the  Einasto model. 

 As to the sub-haloes contribution, again the APS is much milder in the case of the cored MS profile
 than the Einasto one.
At $l$=1000, the APS for the two models differs by more than two orders of magnitude. 
 In the case of Einasto profile, 
 the  emission from the clumps is very anisotropic 
 and similar to the emission of a point-source 
 population, while in the case of the cored MS profile the 
 $\gamma$-ray flux from each halo is more smoothly distributed over its radial dimension. 
 The result is that the APS of the sub-structures for the Einasto profile is more 
 concentrated in the center, i.e. higher in normalization, with respect to the MS one since the clumps, appearing more as point-like, 
 inject more power at all scales. 
This can also be understood by inspecting the all-sky $\gamma$-ray maps (see Figs.~\ref{fig:map}): comparing the Einasto and MS 
parameterization it is clear that the emission from the MS profile is more isotropic on the sky than the Einasto one.
The sub-haloes APS trend for the Einasto and MS profiles is very similar up to $l\simeq$ 100, the former being stronger by a factor 15-20. 
Both curves grow proportionally to $l^2$ as typically expected for a population of point-like sources.
For higher multipoles, the APS starts to flatten because the central part of the sub-haloes starts to be resolved \citep{2009PhRvD..80b3520A}. 
This property is striking for the MS case, for which the sub-haloes $C_l$ spectrum flattens around $l\simeq$ 400 and then decreases significantly. 
Indeed the core of the sub-structures is, on average, $\sim 1.5$ kpc which corresponds to a multipole $\ell \sim 300$ for a source at $\sim 200$ kpc as it is the average distance of the clumps in the g15784 simulation.
The flattening for the Einasto case is much milder and occurs at smaller scales because within this profile the core is less pronounced.

As clear from Fig. \ref{fig:cl}, the APS yielded by the Einasto profile is dominated by the smooth halo up to $l\simeq$ 1000. At variance, 
the same galactic halo interpreted in terms of the MS radial profile yields an APS for the sub-haloes which dominates over the smooth halo
for $\ell\gsim$ 250. In principle, future observations of the shape of the APS ascribable to DM, 
will allow to explore the distribution of galactic DM at scales smaller than the resolution of $N$-body simulations.
The study of high-multipoles anisotropies 
- achievable by the next generation of Cherenkov telescopes such as CTA \citep{2014JCAP...01..049R} -
might help in the debate about the real shape of the DM distribution in the center of the galaxies, 
and in particular of the Milky Way. 
We also notice that the computation of the APS relies on the full $\gamma$-ray sky maps and, thus, does not mask any part of the sky that 
would be required in order to compare our prediction with the {\it Fermi}-LAT results. 

\begin{figure}
\includegraphics[width=0.5\textwidth]{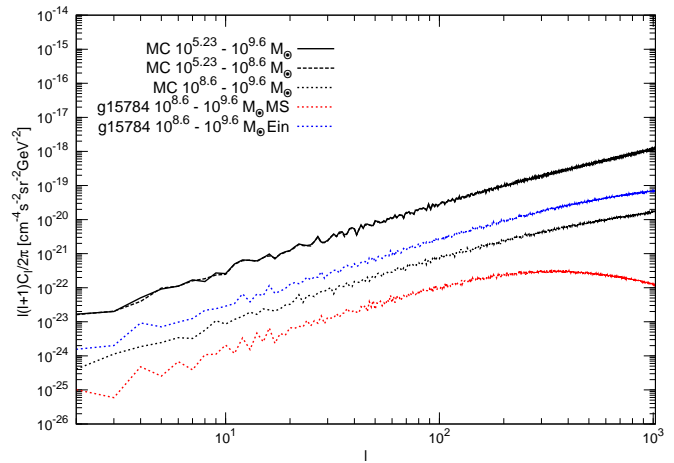}
\caption{Intensity APS computed for the sub-haloes resolved by the g15784 simulation, when the DM density profile 
is interpreted with an Einasto (upper blues solid line) or a MS profile (lower red solid line). 
We also plot the results from a Monte Carlo (MC) realization based on Aquarius Aq-A-1 simulation, 
for sub-haloes having masses from $10^{5.23} - 10^{9.6} M_{\odot}$  (black solid line),    $10^{5.23} - 10^{8.6} M_{\odot}$  (black dashed line)
and  $10^{8.6} - 10^{9.6} M_{\odot}$ (black dotted line).}
\label{fig:cl_MC}
\end{figure}

In order to verify the role of the orientation of the sub-haloes on the APS of the g15784 simulation, 
we have  generated about 850 Monte Carlo realisations in which we have randomly assigned 
the latitude and longitude of each original sub-halo, while keeping fixed the distance. 
The corresponding 90\% confidence level uncertainty band for the Einasto parameterization of the sub-haloes, 
is shown by the grey band in  Fig.~\ref{fig:cl} (we have verified that the simulated 
$C_{\ell}$ distribute normally). 
The fact that the APS of the g15784 simulation stands in the upper edge of the band 
is somewhat expected, since sub-haloes in a DM only 
simulation are usually distributed anisotropically and are preferentially located along the major axes of 
the triaxial mass distributions of their hosts (e.g Zentner et al. 2005).
By randomising their positions we tend to go towards a more isotropic distribution that differs from
the original simulated one.

Finally, we have inspected the effect of sub-haloes smaller that the ones obtained in the present cosmological simulation. 
Fig. \ref{fig:cl_MC} depicts the APS computed for the set of sub-haloes resolved by the g15784 simulation
(same as in Fig. \ref{fig:cl}), again interpreted 
both within Einasto and MS DM profiles. In addition,  we also report the APS generated by a realisation of our 
Monte Carlo simulation based on the Aquarius Aq-A-1 results \citep{aquarius}. For this purpose, we used the spatial, mass and concentration distributions for the clumps population given by \cite{2011PhRvD..83b3518P}, and we assume the DM profile in both the main halo and the sub-haloes to follow the Einasto parameterization with $\alpha_E=0.18$.
For an easier comparison, we show the APS for different mass ranges: $10^{5.23} - 10^{9.6} M_{\odot}$, $10^{5.23} - 10^{8.6} M_{\odot}$ and 
 $10^{8.6} - 10^{9.6} M_{\odot}$. The more 
 massive haloes lead to the flattening of the APS at large multipoles, as expected, while the contribution of the sub-structures 
 lighter than $10^{8.6} M_{\odot}$ is slightly more Poisson-like,
and dominates the total APS, which results to be more intense because of this additional component. 
Given the uncertainties in extrapolating the mass-concentration relation beyond the resolution of the simulations 
(\cite{2013arXiv1312.0945L, 2013arXiv1312.1729S}), we decided not to consider masses smaller that the Aq-A-1 resolution ($\sim 10^{5} M_{\odot}$).

In the present analysis we have not included any contribution from DM in extragalactic structures. 
As discussed in \cite{2013MNRAS.429.1529F} (see also \cite{2014arXiv1401.2117S}), the 
contribution from extragalactic DM halos and sub-halos that are not resolved by $N$-body simulations leads 
to about two orders of magnitude 
uncertainty on the predicted level of the extragalactic energy spectrum, 
which may result as  the dominant or the sub-dominant component of the total
 energy spectrum. Similarly, the intensity APS can receive a significant  
 or a negligible contribution from extragalactic (sub)structures.

\section{ Conclusions} 
We have calculated the intensity APS of the $\gamma$-ray flux from DM annihilation in the halo of a Milky Way like galaxy, employing the original 
results from  recent numerical simulations of structure formation, which predict the DM phase space and its clustering into sub-haloes. 
The simulated galactic halo and its sub-haloes can be equally well  interpreted in terms of a peaked Einasto  
as well as an asymptotically cored MS radial dark matter profile. 

We show here that the different parameterizations for the DM density distribution leads to very different predictions for the $\gamma$-ray intensity APS. 
The DM halo and sub-haloes, when interpreted in terms of the peaked Einasto profile, yield much higher APS at small radial scales (high $l$) than the 
cored MS $\rho(r)$.  The two profiles give comparable APS for the main halo only for $l\lsim$ 10, while at $l$=100 the Einasto APS is about 
two orders of magnitude higher than the MS one. At very small scales, $l\simeq$1000, the main halo within 
the MS profile does not contribute any longer to the 
anisotropy of the sky, while the Einasto profile still provides a sizable APS (about eight orders of magnitude above the MS contribution). 
We have also proven that the sub-haloes APS is significantly  lower for the MS case. Indeed, 
 the APS of the sub-structures described by the Einasto profile is higher than the MS one
 since the sub-haloes, appearing as point-like sources, inject more angular power at almost all scales. 

Our results demonstrate that the extrapolation of the radial DM profile down to radii not proven by cosmological simulations is
specially dangerous when dealing with the search for anisotropies in the $\gamma$-ray emission. The results for the APS 
at high multipoles may differ by huge amounts by a mere re-interpretation of the simulated haloes with a different DM radial 
density distribution. Also, depending of the assumed profile, it may occur that the sub-haloes give a peculiar 
signature in the APS or, at variance, that the main halo dominates at all multipoles the $\gamma$-ray emission. In the latter
case, the APS signature for DM annihilating in the galactic halo is significantly weakened. 

As a final comment, we underline the caution in adopting extrapolated DM profiles when dealing with anisotropy searches, 
and emphasize the need for a better knowledge of the distribution of the DM in its clustered structures, 
especially taking into account the possible effects of baryonic matter (e.g. 
\citet{Maccio2012, 2014MNRAS.437..415D}).
On the other hand, $\gamma$-ray anisotropy analysis will turn out to be crucial 
for probing the spatial DM distribution in the galaxy. Indeed, high multipoles 
measurements will probe scales well beyond the simulations' resolution and will help in discriminating the 
DM profile at very small radii.

\section*{Acknowledgments} 
 We acknowledge  E. Borriello for the discussions during the first stages
of this work and H. Zechlin for a careful reading of the manuscript. 
Moreover, we are grateful to the anonymous referee for having suggested us interesting developments of the discussion.
F.C. acknowledges support from the German Research Foundation (DFG) through grant BR 3954/1-1 and, for the latest stage of this work, from the European Research Council through
the ERC starting grant WIMPs Kairos, P.I. G. Bertone.
F.C. is grateful to S. Ando, T. Bringmann, D. Horns and H.S. Zechlin
for useful and stimulating discussions.
V.D.R acknowledges  support from the EU Network grant UNILHC
PITN-GA-2009-237920 (Universidad de Valencia) for the first stages of this
work and from the  EU FP7 ITN INVISIBLES (Marie Curie Actions,
PITN-GA-2011--289442).
A.V.M. acknowledges the support from the 
Sonderforschungsbereich SFB 881 "the Milky Way System" (subproject A1) 
of the German Research Foundation (DFG).
J.H. received funding from the 
European Research Council under the European Union's 
Seventh Framework Programme (FP 7)  ERC Grant Agreement n. [321035].
For the analysis we used the "Amiga Halo Finder" (AHF, 
\url{popia.ft.uam.es/AHF/}) for identifying halos and the {\sc pynbody} 
package (\url{github.com/pynbody/pynbody/}) for their analysis.

\appendix

\bibliographystyle{mn2e}
\bibliography{paper6}

\end{document}